# Extending System Performance Past the Boundaries of Technical Maturity: Human-Agent Teamwork Perspective for Industrial Inspection

Garrick Cabour[1], Élise Ledoux[2] and Samuel Bassetto[1]

[1] Department of Mathematics and Industrial Engineering, Polytechnique Montreal, P.O. Box 6079, Station Centre-ville, Montreal, QC H3C 3A7, Canada
`garrick.cabour@polymtl.ca`

[2] Physical Activity Department, Université du Québec à Montréal, Montréal QC H3C 3P8, Canada

**Abstract.** Cyber-Physical-Social Systems (CPSS) performance for industry 4.0 is highly context-dependent, where three design areas arise: the artifact itself, the human-agent collaboration, and the organizational settings. Current HF&E tools are limited to conceptualize and anticipate future human-agent work situations with a fine-grained perspective. This paper explores how rich insights from work analysis can be translated into formative design patterns that provide finer guidance in conceptualizing the human-agent collaboration and the organizational settings. The current manual work content elicited is disaggregated into functional requirements. Each function is then scrutinized by a multidisciplinary design team that decides its feasibility and nature (autonomy function, human function, or hybrid function). By doing so, we uncover the technical capabilities of the CPSS in comparison with subject-matter experts' work activity. We called this concept technological coverage. The framework thereof allows close collaboration with design stakeholders to define detailed HAT configurations. We then imagined joint activity scenarios based on end-users work activity, the technological capabilities, and the interaction requirements to perform the work. We use a study on technological innovation in the aircraft maintenance domain to illustrate the framework's early phases.

## 1 Introduction

The technical maturity during the system design cycle is generally measured on the technological development of its features, which does not consider the social-organizational-operational environment in which the technology is implemented [1]. Automation is switching towards autonomy, requiring a more holistic/socio-technical approach to appraise the joint Human-Agent performance in process-tasks completion: "*as the system cannot be neatly divorced from the evaluation of the performance of the user, or the performance of the Human-Machine System [or Cyber-Physical-*

*Social System (CPSS)] as a whole*" (p.22) [2]. An intelligent system can perform well in experimental circumstances in terms of accuracy and output sensitivity (technical maturity). However, this validity does not embrace the operators who will use it in an industrial context or joint Human-Agent performance [3], [4]. The joint performance should be anticipated, evaluated, and iterated during the system design cycle as it will influence the overall CPSS performance, acceptance, and integration [2].

Before implementing a technological solution in complex domains, previous research emphasized the need to understand better how work is currently performed to capture the embedded context ecologically [5]. HF&E methods are well suited to elicit the relevant contextual factors that shape the work domain and taskwork settings. Ergonomic Work Analysis (EWA) is a robust analytical framework that elicits relevant content regarding work domain features and end-users' work activity [6]. However, the challenge is how to use the rich data elicited to formalize efficient Human-Agent Teaming (HAT) configurations? The transition between descriptive fieldwork data obtained with EWA and design remains unclear [7]. Bridging this gap requires formative design models that guide artifact design, human-agent collaboration (HA), and the organizational settings in the envisioned world. [8]. Indeed, current HAT taxonomies offer general design guidelines that encompass function allocation, interaction design, etc. However, two main issues emanate from them. First, the analytical layer is "coarse-grained." It does provide requirements and abstracted design patterns for specific system features (communication modes, type of interaction) [9], [10]. Yet, it limits the guidance provided to designers to conceptualize how human and artificial agents will engage together in a joint activity [4]. Second, the traditional methods of assigning human-machine functions do not systematically impregnate the socio-technical context and the work to be done [5]. These generic taxonomies do not provide a clear integration plan of work analysis, whereas the performance of CPSSs is highly context-dependent [4]. A fine-grained design perspective should adequately conceptualize the HA joint activity to deliver optimal HAT configurations that enhance the overall performance.

## 2   Methods

This research is part of a partnership research project aiming to develop an automated cell to inspect aircraft parts in the aeronautics sector. We present the basis of a framework to show how to provide useful guidance and actionable design patterns from fieldwork inquiries to design a CPSS (that encompasses a HAT perspective).

### 2.1   Research Design

This study is grounded within the *Ergonomic Work Analysis* (EWA) and *Knowledge Engineering* (KE). We conducted qualitative fieldwork to obtain rich insights into the industrial inspectors' work activity: information intake mechanisms, meaning-making and decision processes (steps, variables, assessment of decision outcomes), the knowledge required, rules and strategies applied.

## 2.2 Data collection and Interpretation

We realized 11 on-site observations with concurrent verbalization, 26 mixed-methods interviews (semi-directed, retrospection with work activity content, self-confrontation), and six operational experiments (*microworld*) for a total of 43 hours spent with 12 inspectors. We deployed several units of analysis with different layers of granularity to elicit conceptual and tacit knowledge from the head of subject-matter experts. A "scaffolding" construction of the expert's mental models related to their work context is realized by triangulating data collection methods (fieldwork inquiries and documentation analysis) during nominal/off-nominal working conditions. A detailed description of the fieldwork methodology deployed, from activity analysis to empirical modeling, can be found in [11].

We interpreted the results with a combination of EWA and KE. EWA enables a systemic perspective of end-users work activity, linking the "*what are they doing?*" with "*why are they doing it in such a way?*", i.e., identifying and understanding how the contextual factors shape the work of inspectors. Several iterative steps were carried out with experts to validate the accuracy and reliability of the data collected in the field. We then used KE to identify and capitalize the knowledge captured through *in situ* analyses. The descriptive knowledge models developed can be found in [11].

## 2.3 Early Phases of Cyber-Physical-Social System Design

**From induction to deduction.** We deduced functional requirements that the HAT should execute from the work activity content elicited (e.g., meaning-making and decision-making tasks). For example, in the task "*the inspector detects a nick on the upper area of the part*," at least three functional specifications emerged:
1. The system must detect the defects
2. The system must classify defects type (nick)
3. The system must map and recognize the different area of the part

**Determination of technological coverage.** Then, we introduced the concept of *technological coverage*. Technological coverage represents the functional capabilities of the technical system being design. Instead of focusing on the system components' technical maturity (e.g., hardware equipment's: cameras, sensors), the emphasis is placed on the different tasks to be accomplished for the targeted work activity to endorse a bottom-up approach. Based on the functional description of the current work system, we determined with the design engineers (4 *focus groups* of one hour via videoconference applications) what functions can we automate. For each function, we identified the possible configurations that could be endorsed:
1. Fully automated (autonomy function)
2. Partially automated (hybrid function)
    a. Requiring manual data entry
    b. Not requiring manual data entry
3. Not automated (human function)

**Fine-grained HAT hypotheses.** The next step was to define each function's configuration, the content of each entity's work and the requirements for human-agent interaction. We took a slight switch from traditional function allocation methods ("who does what"). Instead of solely segmenting the human work content and the artificial agent work content, we identified the *inter-functional dependencies* [12] or *interdependencies* [4] that allocation decisions would generate on the other party. In other words, if the agent is in charge of Function A, what does it means for the operators? Based on technological capabilities, in which function/situation does the agent need human intervention (e.g., manual input or performance monitoring) during real-time process tasks execution? Is the output sensible and must be validated by human operators? What information does the operator/agent need to know to continue the sequence of operations?

The work analysis performed in the earliest phases was essential to identify the constraints (or work determinants) that restrict which functions each party can assume or not [5]. As aeronautics is a domain where safety is paramount, the design team decided that the operators would make the final decision on each part inspected.

## 3     Activity Analysis of Industrial Inspection in Aeronautics

Industrial inspection for aircraft maintenance is a highly regulated and safety-critical domain, where human inspectors process multiple sources of knowledge distributed in their environment to diagnose a component's condition. This diagnosis compares the component's state with existing standards that specify the acceptance/refusal criteria and rules to be applied. Inspectors are the central pillar of a decision-making ecosystem that involved other workers (operators, engineers, and technical representatives) with whom they interact to diagnose and repair service-run components correctly. Their work can be broken down as follows (adapted from [13]):

1. Work preparation: inspector sets up the workstation, collect inspection equipment, tasks aids, and aircraft parts to inspect
2. Multi-sensorial search: he examines the part using visual, tactile and perceptual-motor senses until he detects an anomaly
3. Diagnosis: he classifies the type of anomaly encountered and measures/estimates its physical characteristics (depth, length, width). Then, he interprets the measures, process the decision-making variables (Table 1) and the rules to apply to decide whether a defect is acceptable as is, needs to be repaired or is unacceptable
4. Execution: he takes the necessary actions following the previous diagnosis. He decides about the overall part's condition and prescribes the set of repair operations required to restore it and conduct computerized tasks (e.g., record part's information in the database). He repeats operations 2 to 5 until no other anomaly is detected.
5. Work completion: he dispatches the item to the corresponding department

**Table 1.** Decision-making variables shaping inspectors' diagnosis*

|   | Decision-making variables (DMV) | Explanation | Importance |
|---|---|---|---|
| 1 | Depth, width, length, circumference of a defect (physical characteristics) | Tolerance thresholds (TT) are based on dimensional measures | Very high |
| 2 | Physical characteristics borderline with the tolerance threshold | Human judgment under uncertainty. Will be removed with automatic measurement | Very high |
| 3 | Family of defect (defect identification) | TT differ according to the types of defects | High |
| 4 | Blending/dressing restrictions | Some repairs operation can only be applied once or twice; some areas cannot be blended etc. | Very high |
| 5 | Area of defect's occurrence |  | Very high |
| 5.1 | Defect occurrence on critical area (attachment zone) | Material removal on critical areas can engender harmful stress and constraints for adjacent parts | Very high |
| 5.2 | Defect occurrence on unregistered area ("gray area") | If a defect appears in an ill-defined zone, it must be rejected. | Very high |
| 5.3 | Defect occurrence on repaired area (e.g., material thickness with defect removal) | Repairing a defect removes parent material. The thickness of a part must not fall below a certain threshold. | Very high |
| … | … | … | … |

\* These decision-variables were first compiled by the analyst from the raw and processed data collected. He then asked three inspectors to range them by order of importance. We present here only five variables out of 18.

Operations 3 and 4 intertwine the most critical aspects of inspectors' work. During these meaning-making and decision-making steps, inspectors collect, process, and apply relevant knowledge/criteria from their cognitive environment (paperwork, computerized procedures, inspection aids, vibro-engraved information on aircraft part, communication with colleagues, and experiential knowledge) to inspect and decide about each defect's state. We have grouped under the heading of "decision-making variables" all the criteria and factors that modulate inspectors' diagnosis process (Table 1). The next section shows how we translated work activity content into formative insights for the CPSS design.

## 4 Formative Patterns for HAT Configurations

This paper presents how we use descriptive analysis in the on-going research project to configure efficient HAT that takes both entities' advantages. For interested readers, we present how the activity analyses were formalized into design patterns to support software development in [11].

The previous section was inductive, where we gathered relevant knowledge on inspectors' work activity. Here, we switched to a deductive perspective to use the fieldwork analysis in the process design. We first have extracted from each task (either physical, cognitive, or perceptual) decision variables and domain constraints, functional requirements that the technical system must execute in a fully autonomous scenario (Table 2, 1$^{st}$ column). By doing so, we disaggregated each function into sub-function to detail the overall work content and subsequently verified the *technological coverage* for each of them (only the "parent" functions are shown in Table 2). We started to examine whether some functions would disappear, be added, or be modified in the envisioned HAT situations during this stage. Based on technological coverage, the work determinants, and the taskwork requirements, we defined the types of function (autonomy, human, or hybrid) with the HAT teamwork hypotheses' associated work content (Table 2, 4$^{th}$ column).

**Table 2.** Formative table of HAT configurations with technological coverage and work content

| Function | Technological coverage | Function Type | Human-Agent Teamwork Hypotheses |
|---|---|---|---|
| I. Work preparation | | | |
| 1. Collect and process relevant data and information on equipment | Yes (limited extend) | Hybrid function with manual input | System: know the areas of the part inspected and the tolerance threshold associated (TT) Inspectors: manual input of the Serial Number and Part Number, as the system is not able to recognize vibro-engraved characters on the aircraft components |
| 2. Assess the quantity of repairs operation allowed by areas (DMV #4 - blending/polishing restrictions) | Yes | Autonomy function | System: calibrated to understand how much repair is allowed for each registered part of the aircraft component. Whenever a repair operation is required, the system display on-screen the remaining number of repairs authorized, inspectors then validate it. |
| 3. Check correspondence between the information on paperwork, the physical part and computer procedures | No | Human function | Inspectors: continue to double-check the exactness of information before inserting the parts in the automated cell |
| II. Multi-sensorial search | | | |
| 4. Detect eliminatory defects (cracks, burnt zone) | Yes (limited extend) | Hybrid function without manual input | The system can detect major defects; however, it can struggle to classify the defect as significant and eliminatory. The system warns the human operator of a significant defect with the associated zone, and he continues the rest of the operation. |
| 5. Detect minor defects | Yes (limited extend) | Hybrid function without manual input | The system can detect almost all types of defects (mick, dent, scratch, pit, corrosion) but can have difficulty with complex combinations (erosion on the part's edges). Inspectors can check these known areas of difficulties, as they are easily perceptible to them. |
| III. Diagnosis | | | |
| 6. Classify defect (DMV #3 – Types of defect) | Yes (limited extend) | Hybrid function with manual input | The system is not able to classify all defects. It will have to display an alert for "unknown" defects if necessary. The inspector will then decide if classification is required (some types of defects have the same tolerances threshold and others do not). |
| 7. Defect measurement (DMV #1 – Physical characteristics) | Yes | Autonomy function | System: can measure the physical characteristics of a defect and display the results on the screen if required by the inspector. |
| … | … | … | … |

## 5      Discussion and Conclusion

The literature review and practitioners' feedback showed that existing HF&E tools lack fine-grained formative design models to support the engineering of highly innovative and interactive technologies. In this paper, we have presented the preliminary results of some of the key aspects that our design team prospect in designing a Cyber-Physical-Social System. We show how to construe work content elicited from activity analysis into formative patterns that guide HAT configurations in the envisioned situations. By translating descriptive fieldwork data into functional specifications, we can compare system capabilities with inspectors' expertise in cross-functional collaboration with automation engineers. We called this step *technological coverage*. Having in mind the strength and weakness of both entities – plus a detailed vision of the socio-technical context of inspection – the design team can define and choose among the best HAT configurations possible that enhance the overall CPSS performance. We believe that this type of fine-grained framework allows early anticipation of the changing work system: people role, work organization (work added, reassigned, and removed for affected stakeholders), and *inter-functional* needs/requirements for the coordination of the joint activity (Table 2).

However, the nature of these scenarios is prescriptive by nature. They uncover software requirements, HAT requirements, and design decisions. Design assumptions should be validated by eliciting the future joint activity through participatory workshops with relevant stakeholders [14]. To proceed with the design process, we plan to animate work simulations with different materials (interactive workflow diagram and hi-fi prototypes in this project) to elicit and evaluate the Human-Agent joint activity: the orchestration of the activity, the fluid coordination of shared human-agent functions, the difficulties encountered (types and causes) and alternatives solutions or "prevented actions" and the other potential work configurations.

**Acknowledgments.** This research is supported by the *Consortium for Research and Innovation in Aerospace in Québec,* funded by Mitacs accelerate program (contract IT11797).